\documentclass [12pt]{article}
\usepackage{graphicx}
\usepackage{amsmath}
\usepackage[psamsfonts]{amssymb}

\begin{document}
\footnotesize{\noindent
{\footnotesize{\noindent{\textit{International Journal of Simulation and Process Modelling}
\textbf{2}, Nos. 1/2, pp. 67-79 (2006) [a special issue on
``Mathematical Modelling and Simulation for Industrial
Applications"; Editors: S. Olariu (USA), B. B. Sanugi (Malaysia)
and S. Salleh (Malaysia)}}}].

\begin{center}
{\LARGE{\bf An Extended Interpretation of the Thermodynamic Theory
Including \\  an Additional Energy Associated \\ with a Decrease in Mass \\}}
\end{center}

\bigskip

\begin{center}
\large \textbf{Volodymyr
Krasnoholovets}$^{1}$  \textbf{and Jean-Louis Tane$^{2}$}
\end{center}

\bigskip
\noindent $^{1}$Institute for Basic Research, 90 East Winds Court,
Palm Harbor, FL 34683, U.S.A., \ {v\_kras@yahoo.com}

\bigskip
\noindent $^{2}$Formerly with the Department of Geology, Joseph
Fourier University, Grenoble, France, \  {TaneJL@aol.com}

\bigskip
\bigskip

\noindent \textbf{Abstract.} Although its practical efficiency is
unquestionable, the thermodynamic tool presents a slight
inconsistency from the theoretical point of view. After exposing
arguments, which explain this opinion, a suggestion is put forward
to solve the problem. It consists of linking the mass-energy
relation $ \pm \Delta {\kern 1pt} m{\kern 1pt} {\kern 1pt} c^{2}$
to the laws of thermodynamics. This combination results in a
clarification of the theory and opens a bridge between
thermodynamics and gravitation. It is shown that the
submicroscopic consideration, which is based on the cellular
concept of the real physical space when the space is treated as
the tessellation lattice, indeed strongly supports the idea of a
variation in mass in thermodynamic processes.

\bigskip



\bigskip\bigskip

\noindent \textbf{Biographical notes:} Dr. Volodymyr
Krasnoholovets is senior scientist at the Institute of Physics,
National Academy of Sciences of Ukraine, and Professor of the
Institute for Basic Research, Florida, USA. His research touches
conventional, fundamental and applied physics. The most
internationally known results were obtained in the physics of
hydrogen-bonded systems and the fundamental physics. In the realm
of the fundamentals, jointly with late Professor Michel Bounias
(Universit\'e d'Avignon, France) they constructed a detailed
mathematical theory of the real physical space that appears as a
tessellation of primary topological balls. The theory has already
received an international recognition, first of all among
mathematicians. The theory strongly supports Krasnoholovets'
submicroscopic mechanics developed in the real space on the
Planck scale $\approx 10^{-35}$ m which is easily transformed to
the abstract formalism of conventional quantum mechanics
constructed on the atomic scale, $\approx 10^{-10}$ m. For more
details see his web site http://www.inerton.kiev.ua.

\medskip
\noindent
Dr. Jean-Louis Tane is a retired scientist whose speciality is geology that he
taught during more than thirty years at Joseph Fourier University, Grenoble,
France, and during three years at the University of Marrakech, Morocco. Most
of his papers deal with geologic topics, particularly the problem of the origin of
spilitic rocks, yet one of his courses concerned the use of thermodynamics in
geochemistry. His interest for physics has been confirmed recently, by the
writing of several papers suggesting a link between thermodynamics and
relativity. They have been published, between 2000 and 2004 in the Journal of
Theoretics.

\bigskip
\bigskip

\section*{1. Statement of the problem}

\hspace*{\parindent} Following typical thermodynamic courses (see,
e.g. Refs. [1-4]), let us consider the experimental context of an
isolated system composed of two parts separated by a mobile
piston. We assume that part 1 contains a gas whose initial
pressure is ${P}_{1}$ while part 2 is a vacuum. If the piston,
previously locked, is released, the gas expands into the whole
system and the thermodynamic interpretation usually given to the
process is based on the two following equations
\begin{equation}
{dU = dQ + dW}, \label{1}
\end{equation}

\begin{equation}
{dU = TdS - PdV},  \label{2}
\end{equation}

\noindent whose significations and mutual relations are
well-known. A more conventional writing of eq. (1) would be ${dU =
\delta Q + \delta W}$, but the simplified and general
transcription adopted here has no influence on the discussion
presented below. Note that the usual reasoning tend to focus
attention on the gas for which it is successively written:

\medskip

\noindent \quad ${dQ} = 0$ (since the gas cannot exchange heat with the
vacuum);

\bigskip

\noindent \quad ${dW} = 0$ (since, by definition, $dW = - P_{\rm ext}
dV$, where $P_{\rm ext}$ is the pressure external to \break
\noindent \quad the gas, i.e. the pressure $P_{\rm ext} = 0$ of the vacuum and $dV$ the change
in volume of the gas);

\bigskip

\noindent \quad ${dU} = 0$ (being the sum ${dQ + dW}$);

\bigskip

\noindent then owing to the equality $dU = 0$ we have $ dS = (P/T)
dV $ where the three quantities \textit{P}, \textit{T} and ${dV}$
are positive and hence ${dS}$ is positive too.

On this basis, it is implicitly admitted that the relation $dU =
0$ is true not only for the gas, but equally for the whole system
defined as isolated. It is precisely upon this point that an
important and interesting problem arises.

To be rapidly convinced of this reality, let us imagine an
experimental context identical to the one just considered, except
that instead of being vacuum, part 2 would contain a gas whose
initial pressure is $P_{2}$. In such a case, applying equation $dW
= - P_{\rm ext }{dV}$ to part 1 and part 2 leads to the doublet of
equations:
$$
{dW}_{1} = {- P}_{2}{\kern 1pt} dV_{1},
$$
$$
{dW}_{2} = {- P}_{1} {dV}_{2}.
$$

Noting that ${dV}_{2} = {- dV}_{1}$, we get for the whole system

\begin{equation}
dW = dV_{1} (P_{1} - {P}_{2}). \label{3}
\end{equation}

Since dV1 is positive when $P_1 > P_2$ and negative when $P_1 < P_2$,
equation (3) shows that $dW$ is always positive  and reaches the
value zero only when ${P}_{1} = {P}_{2}$. In other words, we are
facing the fact that the disappearance of the pressure gradient
within the system is accompanied by an additional energy whose
origin in inside this system.

\medskip

The process examined just before (expansion of a gas in vacuum)
corresponds to the particular case for which ${P}_{2} = 0$ in
equation (3), so that ${dW}$ has a positive value, not a zero
value as previously recalled and classically accepted. This is the
sign that the vacuum plays an important role in the calculation.
When equation ${dW} = {- P}_{\rm ext}{dV}$ is applied to it,
${P}_{\rm ext}$ corresponds to the pressure of the gas and ${dV}$
to the decreasing volume of the vacuum, which can be noted
${dV}_{\rm vacuum}$. Accordingly, $dW_{\rm vacuum}$ has a positive
value and represents at the same time ${dW}_{\rm system}$, given
that ${dW}_{\rm gas}$ is zero.

\medskip

Transposing these results in equation (1), and applying it to the
whole system, we see that the only condition to obtain ${dU}_{\rm
system} = 0$ is that ${dQ}_{\rm system}$ should be negative. Such
a condition is not compatible with one of the fundamental laws of
thermodynamics (sometimes called the law of heat exchange), which
relies on the idea that an exchange of heat between two parts 1
and 2 of an isolated system is described by the relation ${dQ}_{2}
= {- dQ}_{1}$. At the scale of the whole system, this relation
leads to the conclusion
\begin{equation}
dQ = dQ_{1} + dQ_2 = 0. \label{4}
\end{equation}

\noindent where ${dQ}_{1}$ and ${dQ}_{2}$ represent the change in
heat for the gas and the vacuum.

As will be explained further, it is highly probable that the true
value of ${dQ}$ is not zero but positive. At the present stage of
the discussion, combining the idea that ${dQ}$ is not negative
with the fact, already noted, that ${dW}$ is positive provides a
sufficient condition to conclude that ${dU}$ is itself positive.
Hence, there is really something not clear in the usual conception
of thermodynamics, which is based on the idea that the internal
energy change ${dU}$ for an isolated system obeys the double
equation
$$
dU = dQ + dW = 0. \eqno(1b)
$$

\section*{2. A suggestion to solve the problem}

\hspace*{\parindent}
We have seen in the section 1 that the disappearance of the
pressure gradient within an isolated system is accompanied by a
creation of energy inside this system. Previously we [5-9]
preliminary analyzed of the problem and proposed a way to its
solution. Now let us elucidate all the aspects of the problem in
detail.

We can give to this energy created inside the system studied the
designation ${dW}_{\rm int}$, in order to distinguish it from the
energy ${dW}_{\rm ext}$ that (in the more general case of a non
isolated system) corresponds to the exchange of work between the
system and its surroundings. With these designations, ${dW}$ takes
the signification
\begin{equation}
{dW = dW}_{\rm ext} + {dW}_{\rm int}. \label{5}
\end{equation}

The condition ${dW}_{\rm ext} = 0$ means that the system is
isolated (in the sense classically given to this concept in
thermodynamics). The condition ${dW}_{\rm int} = 0$ means that
there is no pressure gradient (i.e. no irreversibility of
pressure) within this system. Contrary to ${dW}_{\rm int}$, which,
as already seen, obeys the relation
\begin{equation}
{dW}_{\rm int} \geq 0. \label{6}
\end{equation}

The value of $dW_{\rm ext}$ can be positive or negative, depending
on whether the system receives work or gives work in its exchanges
with the surroundings.

If the system we are considering is not isolated but only closed,
the work ${dW}$ exchanged with its surroundings (and whose precise
designation is ${dW}_{\rm ext}$) is given, as recalled above, by
the classical expression $dW = - P_{\rm ext} {\kern 1pt}{dW}$. If
we are under conditions of reversibility, i.e. if ${P}_{\rm ext} =
{P}_{\rm int}$, this relation becomes $dW = - P_{\rm int} {dV}$.
Observing that ${P}_{\rm ext}$ can always be written ${P}_{\rm
ext} = {- P}_{\rm int} + {P}_{\rm int} + {P}_{\rm ext}$, we get
the general; expression
\begin{equation}
{dW = - P}_{\rm int} {dV} - {(P}_{\rm ext}{-P}_{\rm int}{)} {dV},
\label{7}
\end{equation}

\noindent which means
\begin{equation}
{dW}_{\rm irr} = {dW}_{\rm rev} + \textit{dW}_{\rm int}, \label{8}
\end{equation}

\noindent that is
\begin{equation}
{dW}_{\rm irr} = {dW}_{\rm ext} + {dW}_{\rm int}. \label{9}
\end{equation}

Each term of eqs. (8) and (9) has the same value and signification
as the corresponding term of equation (7). An important point to
be noted is that ${dW}_{\rm ext} = {dW}_{\rm rev}$.

As concerns the exchanges of heat and their relations with the
concept of entropy, the usual thermodynamic tool is given by the
well-known expression
$$
dS \geq dQ / T_{\rm ext},
$$

\noindent
which can be represented in the form
\begin{equation}
{dS = dQ/T}_{\rm ext} + {dS}_{\rm int}. \label{10}
\end{equation}

To be coherent with the reasoning previously adopted, eq. (10),
which has the dimensionality of entropy, needs to be converted in
an equation having the dimensionality of energy, i.e.
\begin{equation}
{T}_{\rm ext} dS = dQ + T_{\rm ext} dS_{\rm int}. \label {11}
\end{equation}

\noindent whose signification is
\begin{equation}
dQ = dQ_{\rm ext} + {dQ}_{\rm int}, \label{12}
\end{equation}

\noindent that is
\begin{equation}
{dQ}_{\rm irr} = {dQ}_{\rm rev} + {dQ}_{\rm int}. \label {13}
\end{equation}

Each term of eqs. (12) and (13) has the same value and
signification as the corresponding term of eq. (11). As well as we
have observed that ${dW}_{\rm ext} = {dW}_{\rm rev}$ we see here
that ${dQ}_{\rm ext} = {dQ}_{\rm rev}$ but this last equality is
not a novelty. In the usual conception of the thermodynamic
theory, it is an implicit equation whose integration leads to eq.
(21) that will be used further.

In the same manner as we have introduced a difference between
${dW}_{\rm ext}$ and ${dW}_{\rm int}$ in eq. (5) and a difference
between ${dQ}_{\rm ext}$ and ${dQ}_{\rm int}$ in eq. (12), we now
easily conceive that there is a difference between ${dU}_{\rm
ext}$ and ${dU}_{\rm int}$. It can be expressed through the
general equation
\begin{equation}
{dU = dU}_{\rm ext}  + dU_{\rm int}, \label{14}
\end{equation}

\noindent which can be better written (as explained hereafter) in
the form
\begin{equation}
dU^* = dU_{\rm ext}  + dU_{\rm int} \label{15}
\end{equation}

\noindent whose meaning is
\begin{equation}
{dU}_{\rm irr} = {dU}_{\rm rev} + {dU}_{\rm int}. \label{16}
\end{equation}

For reasons of homogeneity with eq. (15), equations (5) and (12)
can be written respectively
$$
dW^* = dW_{\rm ext} + {dW}_{\rm int}
$$

\noindent and
\begin{equation}
dQ^* = dQ_{\rm ext} + dQ_{\rm int}. \label{17}
\end{equation}

In conventional thermodynamics, ${dU}_{\rm ext}$ is generally
noted ${dU}$. It is to avoid the risk of an additional confusion
that the sum ${dU}_{\rm ext} + {dU}_{\rm int}$ has been noted
${dU^*}$ (and not simply ${dU}$) in eq. (15). The conceptual
difficulty which is often encountered in learning (and teaching)
thermodynamics comes from the fact that ${dU}_{\rm int}$ is not
recognized as an energy, so that ${dU^*}$ and ${dU}_{\rm ext}$ are
not differentiated. Consequently, they are expressed through the
same designation ${dU}$ with the drawback evoked above.

We have just recalled that the term usually noted ${dU}$ in
thermodynamics corresponds to the one noted ${dU}_{\rm ext}$ in
eq. (15). Concerning the energetic quantity ${dU}_{\rm int}$ and
given that it is generated within the system, there is no other
solution to explain its origin as that of a correlative decrease
in mass, according to the Einstein mass-energy relation. Since
this relation was not known when the laws of thermodynamics have
been stated, we easily conceive that it was not evident for their
authors to imagine such a possibility, all the more that the
decrease in mass is generally much too small to be experimentally
detectable.

The practical and well-known efficiency of the thermodynamic tool
is evidently due to the fact that although not identified as an
energy, the term ${dU}_{\rm int}$ is nevertheless taken into
account through the concept of the internal component of entropy,
usually noted ${dS}_{\rm int}$. We have to note that being a
physical reality, ${dU}_{\rm int}$ is not easy to eliminate. It is
certainly for this reason that the art of playing thermodynamics
often looks like an acrobatic feat whose objective is to convert
${dU}_{\rm int}$ into ${dS}_{\rm int}$ everywhere it appears. For
instance, at the study of anomalies in water, we [10,11] were
forced to introduce an additional correction to the Gibbs
potential called ``the change of the Gibbs potential associated
with intermolecular interaction (or the configuration potential)"
$G_{\rm in.in.}= (C_{\rm p} - S)\Delta T $; this potential
definitely introduced a correction to the entropy $S$ of the
water system studied.

On the other hand, let us now look at the Einstein mass-energy
relation that classically is presented through the expression $ E
= mc^{{\kern 0.7pt}2}$ where $c$ is the speed of light. In the
differential form its becomes
$$
dE = c^{2} dm,
$$

\noindent which would be preferably written for our case as $dE =
- c^{{\kern 0.7pt} 2 }{dm}$, since ${dE}$ is positive when ${dm}$
is negative and conversely. Transposed in the language of
thermodynamics and taking into account the term ${dU}_{\rm int}$
previously explained, this expression becomes
\begin{equation}
{dU}_{\rm int} = - c^{{\kern 0.7pt} 2} {dm}. \label{18}
\end{equation}

We have seen above that ${dU}_{\rm int}$ is the energy created
within a system, due to the disappearance of an internal gradient
that is of an internal irreversibility. Combining this data with
the observation, evidenced for a long time by scientists, that
every kind of irreversibility generates heat, both the term
${dW}_{\rm int}$ of eq. (5) and the term ${dQ}_{\rm int}$ of eq.
(13) can appear as being heat.

The great difference between the theory suggested here and the
classical one lies in the origin of the heat, which is linked to
irreversibility. In conventional thermodynamics, this heat is
considered as coming from the fact that the energy received by the
system, from its surroundings, is partly transformed into thermal
energy. In our theory, the energetic quantity ${dU}_{\rm int}$ is
considered as resulting from a partial disintegration of mass
inside the system. For a better understanding of this difference,
a practical and numerical example is examined below.

\section*{3. A numerical application of the suggested theory}

\hspace*{\parindent} Let us consider an isolated system composed
of two parts 1 and 2, each of them containing liquid water. We
suppose that the initial temperature ${T}_{1}$ of part 1 is 293 K
(20$^\circ$ C) and that the initial temperature ${T}_{2}$ of part
2 is 333 K (60$^\circ$ C). We know that the evolution of the
system consists of an irreversible exchange of heat between the
two parts, until they reach the same final temperature $T_{\rm
f}$. We also know that\textit{ T}$_{f}$ is given by the relation
\begin{equation}
T_{\rm f} = (C_{1} T_{1} + C_{2}T_{2}) / (C_{1} + C_{2}) \label
{19}
\end{equation}

\noindent where ${C}_{1}$ and $C_{2}$ are the heat capacities of
part 1 and part 2, respectively. To simplify the calculation
without reducing its interest, we can assume that the mass of
water is the same in each part and choose it in such a way that
$C_{1} = {C}_{2} = 1$ J. Since C = 4.18 J for one gram of water,
the mass we have to consider is $1/(4.18) = 0.223$ [g]. Knowing
that liquid water is a condensed form of water, we can also note
that\textit{ C} designates indifferently $C_{v}$ and $C_{\rm p}$
(whose values are quite equal in such a case) and that we make an
acceptable approximation in considering that ${C}$ does not vary
significantly with temperature.

Admitting these considerations, the difference between the classical
interpretation and the one suggested presently is the following.

\subsection*{3.1. The classical interpretation}}

Eq. (19) gives
\begin{equation}
\label{20} {T}_{\rm f} = (1  \times  293 + 1  \times  333)/ (1 +
1) K = 313 {\rm K}.
\end{equation}

From the general relation
\begin{equation}
\Delta  Q = C  \times  (T_{\rm f} - T_{\rm i}) \label{21}
\end{equation}

\noindent where $T_{i}$ means $T_{\rm initial}$ and having, in the
present case, $\Delta U = \Delta Q$ (since $\Delta W = 0$),

\noindent
we obtain

 $\Delta $ \textit{U}$_{1}$ = $\Delta $\textit{Q}$_{1}$ = 1 $ \times $ (313 -
293) J = 20 J,

 $\Delta $ \textit{U}$_{2}$ = $\Delta $\textit{Q}$_{2}$ = 1 $ \times $ (313 -
333) J = - 20 J

\noindent
so that, for the whole system

 $\Delta $ \textit{U =} $\Delta $\textit{Q =} $\Delta $\textit{U}$_{1}$ +
$\Delta $\textit{U}$_{2} $= 0.

Concerning the entropy change, the general equation is

\begin{equation}
\label{eq22} \Delta {\kern 1pt} {\kern 1pt} S =
\int\limits_{T_{1}} ^{T_{2}} {\frac{{C}}{{T}}dT} ,
\end{equation}

\noindent
which gives

 $\Delta S_{1} = \ln (313/293) = 0.066030$ JK$^{-1}$,

 $\Delta  {S}_{2} = \ln (313/333) = - 0.061939 {\kern 2pt} {\rm JK}^{-1}$

\noindent
so that, for the whole system:

 $\Delta {S} = 0.066030 {\kern 2pt}{\rm JK}^{-1} - 0.061939 {\kern 2pt}
 {\rm JK}^{-1}= 0.00409 {\kern 2pt} {\rm JK}^{-1}$.

The important point to keep in mind is that $\Delta U$ is zero,
while $\Delta S$ is positive.

\bigskip

\subsection*{3.2. The new suggested interpretation}

\hspace*{\parindent}
By integration, eq. (11) leads to an expression, which can be written
\begin{equation}
\label{23} T_{\rm ext}^{\ast}\Delta S = \Delta {Q + T}_{\rm
ext}^{\ast}\Delta S_{\rm int},
\end{equation}

\noindent
whose meaning is
$$
\Delta U^{\ast} = \Delta U_{\rm ext} + \Delta U_{\rm int}.
$$

In eq. (23), $T_{\rm ext}^{\ast}$ represents the average external
temperature that can be calculated as follows:

\bigskip

\noindent for part 1, its value is $T_{\rm ext}^{\ast} =
T_2^{\ast} = \Delta Q_2/\Delta S_2 = - 20/( - 0.061939)$ K =
322.89 K,

\noindent for part 2, its value is $T_{\rm ext}^{\ast} =
T_1^{\ast} = \Delta Q_1/\Delta S_1 = 20/(0.066030)$ K = 302.89 K.

\bigskip

Contrary to $T_1$ and $T_2$, which are varying both in space and
in time, $T_1^{\ast}$ and $T_2^{\ast}$ can be seen as
\textbf{space-time parameters}, whose values are constant for the
process we are considering. In a more general way, this
peculiarity appears as a characteristic of $T_{\rm ext}^{\ast}$.

The terms $\Delta Q_1$, $\Delta Q_2$, $\Delta S_1$ and $\Delta
S_2$ having the numerical values already calculated in section
3.1, when applying eq. (23) to part 1 and part 2 we get
successively

\medskip

\noindent \textbf{for part 1:}

\medskip

$\Delta U^{\ast}_1 = T_{\rm ext 2}^{\ast}\Delta S_1 = 322.89
\times  0.066 030$ J = 21.32 J,

\noindent so that:

\medskip

$\Delta U_{\rm int 1} = \Delta U^{\ast}_1 - \Delta U_{\rm ext 1} =
21.32$ J - 20 J = 1.32 J;

\medskip
\noindent \textbf{for part 2:}

\medskip

$\Delta U^{\ast}_2 = T_{\rm ext 1}^{\ast} \Delta S_2 = 302.89
\times (- 0.061939)$ J = - 18.76 J,

\noindent so that:

\medskip

$\Delta U_{\rm int 2} = \Delta U_2^{\ast} - \Delta U_{\rm ext 2} =
- 18.76$ J - (- 20) J = 1.24 J;

\noindent for the whole system:

\medskip

$\Delta U_{\rm int} = \Delta U_{\rm int 1} + \Delta U_{\rm int 2}
= 1.32$ J + 1.24 J = 2.56 J,

\medskip

$\Delta U_{\rm ext} = \Delta U_{\rm ext 1} + \Delta U_{\rm ext 2}
= 20$ J $- {\kern 2pt} 20$ J = 0,

\noindent so that

\medskip

$\Delta U^{\ast} = \Delta U_{\rm ext} + \Delta U_{\rm int}$, i.e.
$\Delta U^{\ast} = 0$ J + 2.56 J = 2.56 J.

\bigskip

The essential point is that the value $\Delta U_{\rm ext} = 0$
corresponds to the fact that the whole system is defined as
isolated.

The value $\Delta U_{\rm int} > 0$ corresponds to the fact that
the evolution of this system (disappearance of an internal
temperature gradient) is accompanied by an increase in energy
whose value is 2.56 J and whose corresponding decrease in mass
(according to the Einstein mass-energy relation) is
\begin{equation}
\label{24} d{\kern 1pt} m = - 2.56{\kern 1pt}/ \left( {3 \times
10^{8}} \right)^{2}\;  {\rm kg} =  - 2.84 \times 10^{ - 17}
{\kern 3pt} {\rm kg}.
\end{equation}

Although this latter value is evidently too small to be detectable
experimentally, it appears as the logic inference of the reasoning adopted.

\bigskip

\section*{4. The concept of free energy in the suggested theory}

\hspace*{\parindent}
The concept of free energy, introduced by Gibbs, is of great
usefulness in the thermodynamic study of chemical reactions. It is
generally noted \textit{G} and defined by the well-known equation
\begin{equation}
\label{25} G = U + PV - TS
\end{equation}

\noindent whose differential form is
\begin{equation}
\label{26} dG = dU + d(PV - TS)
\end{equation}

The point to be careful with when considering $G$ in the light of
the new suggested theory is that ${dU}$, in eq. (26), corresponds
to $dU_{\rm ext}$ that can itself be written indifferently as
\begin{equation}
\label{27} dU_{\rm ext} = T_{\rm int}^{\ast} dS - P_{\rm
int}^{\ast} dV
\end{equation}

\noindent
whose meaning is  $dU_{\rm ext} =  dU_{\rm reversible}$, or as
\begin{equation}\label{28}
 dU_{\rm ext} = T_{\rm ext}^\ast dS  - P_{\rm ext}^\ast
dV + c^2 dm
\end{equation}

\noindent
whose meaning is
$$
dU_{\rm ext} = dU^\ast - dU_{\rm int}, \ \
$$
i.e.
$$
dU_{\rm ext} =dU_{\rm irr} - dU_{\rm int},
$$
so that
$$
dU_{\rm ext}= dU_{\rm rev}. \ \  \qquad
$$

Thus we see that equation (28) is really coherent with eq. (27).

Taking into account these considerations, the whole meaning of eq. (26) can
itself be understood through two equivalent interpretations:

\medskip

\subsection*{4.1. The first interpretation}
\begin{equation}\label{29}
dG = T_{\rm int}^\ast dS - P_{\rm int}^\ast dV + d(PV - TS),
\end{equation}

\noindent
that is
$$
dG = T_{\rm int}^\ast dS - P_{\rm int}^\ast dV +
P_{\rm ext}^\ast dV + VdP_{\rm ext}^\ast  -
T_{\rm ext}^\ast dS - S dT_{\rm ext}
$$

\noindent
where $P_{\rm ext}$ and $T_{\rm ext}^\ast$ are constant (as already
seen for $T_{\rm ext}^\ast $ in the section 3.2), so that
$dT_{\rm ext}^\ast = 0$ and $dP_{\rm ext}^\ast = 0$. Consequently,

\begin{equation}\label{30}
dG = (T_{\rm int}^\ast - T_{\rm ext}^\ast){\kern 1pt} dS + (P_{\rm ext}^\ast
- P_{\rm int}^\ast) {\kern 1pt} dV.
\end{equation}

\subsection*{4.2. The second interpretation}

\hspace*{\parindent}
The second interpretation is equivalent to the first one, but using
$dU_{\rm ext}$ as defined by equation (28)
\begin{equation}\label{31}
dG = T_{\rm ext}^\ast {\kern 1pt} dS - P_{\rm ext}^\ast {\kern 1pt} dV
+c^{2} dm + d(PV - TS),
\end{equation}

\noindent
which (after the same treatment as that already done), is reduced to
\begin{equation}\label{32}
dG = c^{2} dm.
\end{equation}

Both equations are the same but the first transcription gives the formula
to calculate $dG$ (29), while the second shows that $dG$ is correlated to
a change in mass (32).

For a practical example of the use of these equations, let us come back to
the exchange of heat between the two parts of the isolated system previously
considered and to the interpretation presented in the section 3.2.

We are in a situation for which ${dV}$ is negligible, so that eq. (30)
is reduced to
\begin{equation}\label{33}
dG = (T_{\rm int}^\ast - T_{\rm ext}^\ast){\kern 1pt} dS.
\end{equation}

Having noted above that $T_{\rm int}^\ast$ and $T_{\rm ext}^\ast$ are
constants (referring to the process we are considering), the difference
$T_{\rm int}^\ast - T_{\rm ext}^\ast$ is a constant too and,
consequently, the integration of eq. (33) leads to
\begin{equation}\label{34}
 \Delta G = (T_{\rm int}^\ast - T_{\rm ext}^\ast) \Delta
 S.
\end{equation}

By applying eq. (34) to part 1 and part 2 successively, we get

\bigskip

 $\Delta G_{1} = (302.89 - 322.89)  \times  0. 066030 \  {\rm J} = - 1.32 \
{\rm J}$,

 $\Delta G_{2} = (322.89 - 302.89) \times  (- 0.061939) \ {\rm J} = -
1.24 \ {\rm J}$,

\bigskip

\noindent
so that, for the whole system, $\Delta G =\Delta
G_{1} +\Delta G_{2} = - 2.56$ J.

This result being the same (with an inversion of sign) as the one obtained
in the section 3.2, we see that $\Delta G$ is effectively
equivalent to - $\Delta U_{\rm int}$.

In the case of a chemical reaction, the negative value of $\Delta
G$ corresponds to the disappearance, within the system, of a
gradient in chemical potential. This peculiarity is a well-known data of the
classical theory. The complementary information we obtain now is that the
negative value of $\Delta G$ is accompanied by a decrease in mass.

The fact that conventional thermodynamics does not take into account the
decrease in mass, which is linked to a chemical reaction, is probably the
reason why it is not of common use in the study of nuclear reactions, which
are effectively characterized by a change in mass. It seems that this
difficulty disappears when the enlarged conception suggested here is
substituted to the classical one.

After this preliminary consideration based on macroscopic thermodynamics, we
will now examine the problem from the submicroscopic point of view.

\section*{5. Submicroscopic consideration}

\subsubsection*{5.1. Space structure and quantum mechanics}

\hspace*{\parindent}
Let us turn now to a microscopic consideration of a dynamic
system. What has to be proved is the realistic mass exchange
between the system studied and the ambient space. Obviously
orthodox quantum mechanics cannot solve the problem, because it
treats mass of a quantum system as an invariable classical
parameter. At the same time relativistic experiments demonstrate
that the mass is not a constant but changes with the system's
velocity. This means that something is wrong in our comprehension
of the fundamentals, namely: There is no any connection between
conventional quantum physics and relativity.

Very recently a rigorous mathematical theory of the real physical space has
been developed by Bounias and Krasnoholovets [12-14]. The theory shows that
the real space represents a mathematical lattice, called \textit{the}
\textit{tessellattice}, packing with elementary cells whose size can be
estimated as the Planck one, $\sim $ 10$^{-35}$ m. The tessellattice
represents a degenerate space-time, i.e. in this case all cells (in other
words, balls) are degenerated.

A particled cell provides a formalism describing the elementary
particles proposed in Refs. [13,14]. In this respect, mass is
determined by a fractal reduction of volume of a cell, while just
a reduction of volume as in degenerate cells, which was initially
postulated in Refs. [15-17], is not sufficient to provide mass
(because a dimensional increase is a necessary condition [13]).
Accordingly, if V$_{0}$ is the volume of an absolutely free cell,
then the reduction of volume resulting from a fractal concavity is
the following: V$^{\rm part}$ = V$_{0} - V_{\rm frac}$. The mass
\textit{m} of a particled cell is a function of the
fractal-related decrease of the volume of the ball:
\begin{equation}\label{35}
m \propto \frac{{1}}{{V^{\rm part}}} \cdot \left( {{\rm e}_{\nu} -
1} \right)_{{\rm e}_{\nu
> 1}}
\end{equation}

\noindent
where (e) is the Bouligand exponent, and $\left( {e_{\nu}  - 1} \right)$ the
gain in dimensionality given by the fractal iteration; the index $\nu $ at
the Bouligand exponent specifies the possible fractal concavities affecting
the particled ball [13]. The moment of junction allows the formalization of
the topological characteristics of what is called \textit{motion} in a
physical universe. It is the motion that was called by de Broglie as the
major characteristic determining physics. While an identity mapping denotes
an absence of motion, that is a null interval of time, a nonempty moment of
junction stands for the minimal of any time interval. In our sense, there is
any minimal spatial ``point" at all: only instants that at bottom of fact do
not reflect timely features.

Now let us consider the motion of a particled cell in the
tessellattice. Clearly the massive particled cell cannot move
without contacts with ambient cells. Such interaction creates
excitations in the tessellattice called \textit{inertons} [15,16],
because the excitations carry inert properties of the particle and
they represent a substructure of the particle's matter waves. As
has been shown [15,16] inertons spread from the particle up to a
distance
\begin{equation}\label{36}
\Lambda = \lambda \,c/\upsilon
\end{equation}

\noindent where $\lambda $ can be called the free path length of
the particle, or the particle's spatial amplitude, $\upsilon $ is
the particle's velocity and \textit{c} is the velocity of
inertons, which may be associated with the speed of light (if we
deal with the charged system). Then $\Lambda $ plays the role of
the free path length of the particle's cloud of inertons.

A detailed theory of submicroscopic mechanics developed in the real physical
space has been constructed in Refs. [15-19]. The theory has allowed us to
complete clarify fundamental notions of quantum physics; the theory has
introduced short-range action that automatically means the introduction of a
new kind of carriers - inertons, - which thus become carriers of the quantum
mechanical interaction.

It has been argued that a deformation coat, or a crystallite, is
formed around a created particle in the tessellattice space. The
size of the crystallite in which cells are deformed is associated
with the Compton wavelength of the particle $\lambda _{\rm Com} =
h/mc$ the role of the crystallite is to shield the particle from
the degenerate space. The mechanics constructed is exemplified by
elementary excitations of the surrounding space, i.e. inertons,
which accompany the moving particle. The motion of the particle
has been studied starting from the Lagrangian (it is simplified
here)
\begin{equation}\label{37}
L = g_{ij} \dot {X}^{i}\dot {X}^{j} + {\textstyle{{1} \over
{2}}}\sum\limits_{l} {\tilde {g}_{ij}^{\left( {l} \right)} {\kern
1pt} \dot {x}_{\left( {l} \right)}^{i} {\kern 1pt} \dot
{x}_{\left( {l} \right)}^{j}} - \sum\limits_{l} {\frac{{\pi}
}{{\Theta_{l}} }} {\kern 1pt} {\kern 1pt} \sqrt{g_{ik}{\kern 1pt}g_{kj}} \left[
{X^{i}\dot {x}_{\left( {l} \right)}^{j} + \upsilon _{0}^{i} {\kern
1pt} x_{\left( {l} \right)}^{j}} \right].
\end{equation}

Here the first term describes the kinetic energy of the particle, the second
term depicts the motion of the ensemble of inertons and the third term
characterizes the interaction between the particle and its inertons.
$X^{i},\;\,\dot {X}^{i}$ are components of the particle coordinate and
velocity; $x_{\left( {l} \right)}^{i} $ and ${\kern 1pt} \dot {x}_{\left(
{l} \right)}^{i} $ are components of the \textit{l}th inerton coordinate and
velocity; $g_{i{\kern 1pt} j} $ and $\tilde {g}_{i{\kern 1pt} j}^{\left( {l}
\right)} $ are mass tensors of the particle and its \textit{l}th inerton;
$\upsilon {\kern 1pt} _{0}^{i} $ are components of the initial velocity of
the particle (here $i = x,{\kern 1pt} \;y,\;z$); $\pi /\Theta_{l} $ is the
frequency of scattering of the particle by the \textit{l}th inerton.

In the so called relativistic case, when $\upsilon _{0} $ approaches to the
velocity of light \textit{c}, the initial Lagrangian has been chosen in the
classical form
\begin{equation}\label{38}
L = - mc^{2}\sqrt {1 - \upsilon _{0}^{2} /c^{2}}
\end{equation}

\noindent
in which the following transformation has been made
\begin{equation}\label{39}
\upsilon {\kern 1pt} _{0}^{2} \;\; \to {\kern 1pt} {\kern 1pt}
{\kern 1pt} {\kern 1pt} \left[ {g_{i{\kern 1pt} j} {\kern 1pt}
\dot {X}^{i}\dot {X}^{j} + F\left( {X,x,\dot {x},\upsilon _{0}}
\right)} \right]{\kern 1pt} /g
\end{equation}

\noindent where the function \textit{F} is similar to the second
and third terms in expression (39).

The Euler-Lagrange equations written for the particle and
inertons allow one to obtain paths of the particle and its
inertons [15-17]. The solutions to the equations of motion show
that the particle oscillates along its path, namely, the
particle's velocity changes periodically from $\upsilon _{0} $ to
zero during a spatial interval $\lambda $, which exactly
coincides with the de Broglie wavelength of the particle. Thus
the value of $\lambda = h/(m\upsilon _{\kern 0.7pt 0}) $
determines the spatial period of the oscillatory moving particle.

Furthermore the relationship $\lambda = \upsilon _{0} \Theta$
holds, where hereinafter $1/\Theta$ is the frequency of
scattering of the particle by its inerton cloud. A cloud of
inertons oscillates around the particle and the cloud's range
periodically changes from zero to $\Lambda = c{\kern 1pt}\Theta$.
These two relationships result in relation (36).

Thereby the motion of a particle is deterministic and features the de
Broglie's relationships
\begin{equation}\label{40}
E = h\nu ,\quad \quad \lambda = h/(m\upsilon _{0})
\end{equation}

\noindent
where in our case $E = m\upsilon {\kern 1pt} _{0}^{2} /2$ and $m = m_{0}
/\sqrt{ 1 - \upsilon _{0}^{2} /c^{2}} $.

The transition to Schr\"odinger's and Dirac's quantum mechanics is
ensured by the transition to typical initial classical
Hamiltonians, which are used to start the formalisms.
Relationships (40) enable one to derive the Schr\"odinger
equation and in this case the wave $\psi$-function, which is an
abstract characteristic in the conventional formalism, gains in
importance of the field of inertia of the particle. The field of
inertia is located in the range covered by the amplitude of the
particle's inerton cloud $\Lambda $, which signifies that the wave
$\psi$-function formalism cannot be employed beyond this range
limited by the amplitude $\Lambda $.

Besides, it is important to emphasize that in the case of a many particle
system (for instance, a solid) the role of the spatial amplitude $\lambda $
of a particle, i.e. its de Broglie wavelength, plays the amplitude of
oscillations $\delta {\kern 1pt} g$ of the particle that vibrates in the
vicinity of its equilibrium position. Below we will use this fact at the
analysis of the behavior of the particle mass in a many particle system.

\subsection*{5.2. Space structure, microscopic mechanics and the
phenomenon of gravity}

\hspace*{\parindent}
What exactly do inertons, as carriers of the elementary quantum mechanical
interaction, transfer? To describe quantum mechanics, we have needed to know
the inerton's position, velocity, kinetic energy and momentum. In paper [20]
a detailed consideration of the motion of a particle in the tessellattice
has been performed. The study has elucidated processes of emission and
re-absorption of inertons, which shows that inertons transfer fragments of
the particle's mass to the distance $\Lambda $ from the particle and then
the space tessellattice being elastic returns inertons back to the particle.
The value of mass $m{\kern 1pt} _{\rm inert} $ transferred by an inerton has
been studied in Ref. [21]; it has been argued that $m{\kern 1pt} _{\rm inert} $
falls within the range around 10$^{-85}$ to 10$^{-45}$ kg.

The mass of an inerton is specified by its own dynamics [20] and,
therefore, the Lagrangian (37) should be supplemented by
additional terms. Namely, a moving particle and its inerton cloud
metamorphose each half a period, i.e. $\lambda /2$ and $\Lambda
/2$, respectively: The particled cell and ambient cells involved
in the inerton cloud periodically change, due to the periodic
character of the motion of the particle, from the deformed state
to the tension state. In other words, the mass (a local
characteristic of the tessellattice) is periodically converted to
the rugosity of the tessellattice (a collective property of cells
enfolded in the inerton cloud). Therefore, keeping in mind that
the availability of the mass in a particle means the total mass of
the particle jointly with its inerton cloud, the mass part of the
particle's Lagrangian becomes
\begin{equation}
\label{41} \quad L = {\textstyle{{1} \over {2}}}{\kern 2pt}  \dot {\mu} ^{2}
+ {\textstyle{{1} \over {2}}}{\kern 2pt}\dot {\vec {\xi} }^
{{\kern 1pt} 2} - c\mu {\kern 1pt} \nabla \vec {\xi }.
\end{equation}

Here $\mu = m{\kern 1pt} \left( {\vec {r},{\kern 1pt} {\kern 1pt}
{\kern 1pt} t} \right)/m_{0} $ is the current dimensionless mass
of the \{particle + inerton cloud\}-system; $\vec {\xi}  = \vec
{\xi} \left( {\vec {r},{\kern 1pt} {\kern 1pt} {\kern 1pt} t}
\right)$ is the current dimensionless value of the rugosity of the
tessellattice in the range covered by the system. Geometrically
the rugosity $\vec {\xi} $ depicts the state in which the
tessellattice cells in the region covered by the inerton cloud do
not have any volumetric deformation (i.e. the cells become
massless), but are slightly shifted from their equilibrium
positions in the tessellattice along the particle vector velocity
$\vec {\upsilon} $. This characteristic, the rugosity, also
embraces the state of inflation of cells, especially the particled
cell. The moving particle periodically changes its state from
deformed (the mass state) to inflated and shifted (the tension
state); this conversion occurs after the passing the section
$\lambda /2$ beginning with the initial starting point that has to
be specified by the mass state. In physical language, the rugosity
manifests itself as the tension of space. The last term in
expression (41) introduces the conversion of the mass
state to the tension state where \textit{c} is the rate of this
conversion, i.e. the velocity of inertons.

Proceeding to the Euler-Lagrange equations for variables $\mu $ and $\vec
{\xi} $ we have to use them in the form

\[
\partial /\partial {\kern 1pt} {\kern 1pt} t\left( {\partial L/\partial
{\kern 1pt} \dot {q}} \right) - \delta {\kern 1pt} L/\delta {\kern 1pt} q =
0
\]

\noindent
where $\delta {\kern 1pt} L/\delta {\kern 1pt} q$ is the functional
derivative

\[
\frac{{\delta {\kern 1pt} L}}{{\delta {\kern 1pt} q}} = \frac{{\partial
L}}{{\partial {\kern 1pt} q}} - \frac{{\partial} }{{\partial {\kern 1pt}
x}}\frac{{\partial L}}{{\partial {\kern 1pt} {\kern 1pt} \left( {\partial
{\kern 1pt} q/\partial {\kern 1pt} x} \right)}} - \frac{{\partial
}}{{\partial {\kern 1pt} y}}\frac{{\partial L}}{{\partial {\kern 1pt} {\kern
1pt} \left( {\partial {\kern 1pt} q/\partial {\kern 1pt} y} \right)}} -
\frac{{\partial} }{{\partial {\kern 1pt} z}}\frac{{\partial L}}{{\partial
{\kern 1pt} {\kern 1pt} \left( {\partial {\kern 1pt} q/\partial {\kern 1pt}
z} \right)}}.
\]

The equations for $m$ and $\vec {\xi} $ obtained from eq.
(41) are the following
\begin{equation}
\label{42}
\partial ^{2}\mu /\partial {\kern 1pt} {\kern 1pt} t^{2} - c{\kern 1pt}\nabla \dot
{\vec {\xi} } = 0,
\end{equation}
\begin{equation}
\label{43}
\partial ^{2}\vec {\xi} /\partial {\kern 1pt} {\kern 1pt} t^{2} - c{\kern 1pt}\nabla
\dot {\mu}  = 0
\end{equation}
The availability of the radial symmetry allows the solutions to
equations (42) and (43) in the form of standing
spherical waves [20], which exhibit the dependence $1/r$; in
particular, the solution for the mass is [20]
\begin{equation}
\label{44} m{\kern 1pt} {\kern 1pt} \left( {r,{\kern 1pt} {\kern
1pt} {\kern 1pt} t} \right) = m_{0} \frac{{l_{\rm Planck}} }{{r}}
\Big|  {\kern 1pt} { \cos\left( {\frac{{\pi {\kern 1pt}
r}}{{2\Lambda} }} \right)} \Big| {\kern 1pt} {\kern 1pt} \Big|
{\kern 1pt} {\cos\left( {\frac{{\pi {\kern 1pt} t}}{{2\tilde
{T}}}} \right)} \Big|.
\end{equation}

The solution for the rugosity/tension $\xi $, which in the case of
the radial symmetry becomes a scalar variable, is
\begin{equation}
\label{45} \xi \left( {r,{\kern 1pt} {\kern 1pt} {\kern 1pt} t}
\right) = \xi _{0} \frac{{\Lambda} }{{r}} \Big| {\kern 1pt} {
\sin\left( {\frac{{\pi {\kern 1pt} r}}{{2{\kern 1pt}\Lambda} }}
\right)} \Big| {\kern 1pt} \Big| {\kern 1pt}  { \sin\left(
{\frac{{\pi {\kern 1pt} t}}{{2{\kern 1pt}\Theta}}} \right)} \Big|.
\end{equation}

In expressions (44) and (45) the variable
\textit{r} satisfies the inequalities $l_{\rm Planck} \le r
\le \Lambda $.

Newton's gravitational law immediately follows from expression
(44). Indeed, at $r < < \Lambda $, $t > > \Theta$ and $\upsilon
^{2}/c^{2} < < 1$ we obtain
\begin{equation}
\label{46} m{\kern 1pt} {\kern 1pt} \left( {r} \right) \cong
l_{\rm Planck} {\kern 2pt} m_{0} /r.
\end{equation}

This relation shows the degree of deformation of the space around the
particle. Then multiplying both parts of expression (46) by a factor $ -
G/l_{\rm Planck} $ where $G$ is the Newton constant of gravitation, we obtain
the so-called potential gravitational energy, or Newton's gravitational
potential, of the object in question
\begin{equation}
\label{47}
U{\kern 1pt} {\kern 1pt} \left( {r} \right) = - G{\kern 1pt} m_{0} /r
\end{equation}

\noindent
where the Newton constant simple plays the role of a dimensional
coefficient.

\subsection*{5.3. The mass of a particle in a many particle system}

\hspace*{\parindent}
So far we have treated the behavior of a stationary moving quantum system.
However, if the system in question consists of a huge number of particles
(e.g. a gas, a solid and so on), the Lagrangians (37) and (41) must be
complemented by terms including the interaction between particles. Let us
focus on the modified Lagrangian (41), which should arrive us exactly at the
changes in the behavior of the system's mass revealed at the aforementioned
thermodynamic analysis,
\begin{equation}
\label{48} L_{\rm mass} = \sum\limits_{\vec {n}} {\left(
{{\textstyle{{1} \over {2}}}{\kern 1pt} {{\dot {\mu}} _{\vec
{n}}}^{{\kern 1pt} 2} + {\textstyle{{1} \over {2}}} \dot {\vec
{\xi}} _{\vec {n}}^{{\kern 2pt} 2} - c{\kern 1pt} \dot {\mu}
_{\vec {n}} \nabla \vec {\xi} _{\vec {n}} - {\textstyle{{\pi}
\over {2}}}\sum\limits_{\vec {n}{\kern 1pt} {\kern 1pt} '} {\chi
_{\vec {n}{\kern 1pt} \vec {n}{\kern 1pt} '}} {\kern 1pt} \dot
{\mu} _{{\kern 1pt} \vec {n}} {\kern 1pt} {\kern 1pt} \vec {\xi}
_{\vec {n}'}}  \right)} .
\end{equation}

The parameter $\chi _{\vec {n}{\kern 1pt} \vec {n}{\kern 1pt} '}
$ throws into engagement the mass of one particle described by the radius
vector $\vec {n}$ with the tension of the tessellattice generated by the
other particle depicted by the radius vector ${\vec {n}}'$; in other words,
$\chi _{\vec {n}{\kern 1pt} \vec {n}{\kern 1pt} {\kern 1pt} '} $ is the
scattering frequency of the rate of mass $\dot {\mu} {\kern 1pt} _{\vec {n}}
$ of the $\vec {n}$-particle by the non-homogeneity of space caused by the
rugosity $\vec {\xi} _{{\vec {n}}'} $ of the ${\vec {n}}{\kern 1pt} '$-particle.

The Euler-Lagrange equations of motion constructed on the basis of the
Lagrangian (48) are the following
\begin{equation}
\label{49}
\ddot {\mu} _{{\kern 1pt} \vec {n}} - c\nabla \dot {\vec {\xi} }_{\vec {n}}
- {\textstyle{{\pi}  \over {2}}}\sum\limits_{{\vec {n}}'} {\chi _{\vec
{n}{\kern 1pt} {\kern 1pt} \vec {n}{\kern 1pt} {\kern 1pt} '} {\kern 1pt}
\dot {\vec {\xi} }_{{\kern 1pt} \vec {n}{\kern 1pt} {\kern 1pt} '}}  = 0,
\end{equation}
\begin{equation}
\label{50}
\ddot {\vec {\xi} }_{\vec {n}} - c\nabla \dot {\mu} {\kern 1pt} _{\vec {n}}
{\kern 1pt} {\kern 1pt} + {\kern 1pt} {\kern 1pt} {\kern 1pt} {\kern 1pt}
{\textstyle{{\pi}  \over {2}}}\sum\limits_{\vec {n}{\kern 1pt} {\kern 1pt}
'} {\chi _{\vec {n}{\kern 1pt} {\kern 1pt} \vec {n}{\kern 1pt} {\kern 1pt}
'} {\kern 1pt} \dot {\mu} _{{\kern 1pt} \vec {n}{\kern 1pt} {\kern 1pt} '}}
= 0.
\end{equation}

Equations (49) and (50) allow us to derive the following equation describing
the behavior of the mass
\begin{eqnarray}\label{51}
\ddot {\mu} _{{\kern 1pt} \vec {n}} &-& c^{2{\kern 1pt}} \nabla _{\vec
{n}}^{2} {\kern 1pt} {\kern 1pt} \mu _{\vec {n}} {\kern 2pt}+ {\kern 2pt}
{\textstyle{{\pi}  \over
{2}}}{\kern 1pt} {\kern 1pt} {\kern 1pt} c{\kern 1pt} {\kern 1pt} \nabla
_{\vec n}  {\kern 3pt} \mu _{\vec {n}} \sum\limits_{\vec n^{{\kern 0.5pt} \prime}}
{\chi _{{\vec n}{\kern 1pt} {\vec n}^{{\kern 1pt}\prime} {\kern 2pt}}}  - {\kern 2pt}
{\textstyle{{\pi}  \over {2}}}{\kern 1pt} {\kern 1pt} C{\kern
1pt} \sum\limits_{\vec {n}^{{\kern 1pt} \prime}} {\chi _{\vec {n}{\kern 1pt} \vec
{n}'} {\kern 1pt}}     \nonumber   \\
&-& {\textstyle{{\pi}  \over {2}}}{\kern 1pt} {\kern 1pt} {\kern 1pt} c{\kern
1pt} {\kern 1pt} \sum\limits_{{\vec {n} {\kern 0.4pt}'}} {\chi _{\vec {n}{\kern 1pt} \vec
{n}{\kern 0.5pt}  '} }  \nabla _{\vec {n}{\kern 1pt} '}
{\kern 1pt} {\kern 1pt} \mu _{{\kern 1pt} \vec {n}{\kern 1pt} {\kern 1pt} '}
{\kern 2pt} + {\kern 2pt}\left( {{\textstyle{{\pi}  \over {2}}}} \right)^
{2}\sum\limits_{{\vec
{n}}',\,{\kern 1pt} {\vec {n}}^{{\kern 1pt} \prime\prime }}
{\chi _{\vec {n}{\kern 1pt} \vec {n}{\kern 1pt} '} \chi _
{\vec {n}'{\kern 1pt} \vec {n}{\kern 1pt} '{\kern 0.4pt} '}}
 {\kern 1pt} {\kern 1pt} \mu _{{\kern 1pt}
\vec {n}{\kern 0.5pt} '{\kern 0.5pt} '}{\kern 2pt}  = {\kern 2pt} 0
\end{eqnarray}

\noindent
where \textit{C} is a constant. In the first crude approximation,
eq. (51) can be simplified to
\begin{equation}
\label{52}
\ddot {\mu} _{{\kern 1pt} \vec {n}} - c^{2}\Delta {\kern 1pt} \mu _{{\kern
1pt} {\kern 1pt} \vec {n}} - c^{2}\Delta {\kern 1pt} \mu _{{\kern 1pt}
{\kern 1pt} \vec {n}} + {\textstyle{{\pi}  \over {2}}}\chi ^{2}\left( {n}
\right){\kern 1pt} {\kern 1pt} \mu _{{\kern 1pt} {\kern 1pt} \vec {n}}
{\kern 2pt} \approx C_{\vec {n}},
\end{equation}

\noindent
where $\Delta $ is the Laplace operator, $\chi ^{2}\left( {n} \right) =
\sum\nolimits_{{\vec n}{\kern 0.5pt}',{\kern 3pt}  {\vec n}{{\kern
0.6pt} ''}} {\chi _{\vec {n}{\kern 1pt} \vec {n} {\kern 0.5pt} '}
{\kern 1pt}\chi _{\vec {n}'{\kern 0.5pt} \vec {n}{\kern 0.5pt}
'{\kern 0.4pt} '}}$ and $C_{\vec {n}} = C\sum\nolimits_{{\kern 1pt}
 \vec {n} {\kern 0.7pt} '} \chi _{{\vec
n}{\kern 2pt}  {\vec n}{\kern 0.5pt}'}$;  below we
put $C = 0$, because $C \ne 0$ will introduce in the solution a term that
describes an incessant inflation/disappearance of the total mass of the
system of particles in question. Thus setting $C_{\vec {n}} = 0$ in eq. (52)
and allowing the periodic solution in time, equation (52) is reduced to
\begin{equation}
\label{53}
\Delta {\kern 1pt} \mu _{\vec {n}} - \frac{{\omega _{\vec {n}}^{2} - \left(
{{\textstyle{{\pi}  \over {2}}}} \right)^{2}\chi ^{2}\left( {n}
\right)}}{{c^{2}}}{\kern 2pt}\mu _{\vec {n}} = 0.
\end{equation}

On the assumption of the radial symmetry of each particle, the solution to
equation (53) becomes similar to expression (44), namely
\begin{equation}
\label{54}
m{\kern 1pt} _{\vec {n}} \left( {r,{\kern 1pt} {\kern 1pt} t} \right)
= \frac{{l_{\rm Planck}} }{{r}}{\kern 1pt} m_{0} {\kern 1pt} \left|
{\cos\left( {K_{\vec {n}}{\kern 2pt} r} \right)} \right|{\kern
1pt} \left| {\cos\left( {\Omega _{\vec {n}} {\kern 2pt} t}
\right)} \right|
\end{equation}

\noindent
where we come back to the dimension mass variable and denote the proper
cycle frequency of the particle oscillation and the corresponding wave
vector and the cyclic frequency as follows
\begin{equation}
\label{55}
K_{\vec {n}} = k_{\vec {n}} {\kern 2pt} \sqrt {1 - \chi ^{2}\left( {\vec {n}}
\right){\kern 1pt} {\kern 1pt} \Theta_{\vec {n}}^{2}}, \quad \quad
\Omega_{\vec {n}} = \omega_{\vec {n}} {\kern 1pt} \sqrt {1 - \chi ^{2}\left( {\vec {n}}
\right){\kern 1pt} {\kern 1pt} \Theta_{\vec {n}}^{2}}.
\end{equation}
Here we designated $k_{\vec {n}} = \pi /2\Lambda _{\vec {n}}$ and
$\omega _{\vec {n}} = \pi /2\Theta_{\vec {n}}$.

Thus equations (52) and (53) show that the mass of particle induces
the mass field $m{\kern 1pt} {\kern 1pt} \left( {r,\,\,t} \right)$ in the ambient
space, which obeys the law of standing spherical wave (54). This mass
distribution generates the tension of space $\xi \left( {r,{\kern 1pt}
{\kern 1pt} {\kern 1pt} {\kern 1pt} {\kern 1pt} t} \right)$, which also
adheres the same law of standing spherical wave (45), though oscillates in
the counter-phase.

\subsection*{5.4. An additional mass}

\hspace*{\parindent}
Solution (54) shows that the union of particles in a many particle system
lowers their energy, because $\hbar \tilde {\omega}  < \hbar \omega $
(hereinafter we omit the index $\vec {n}$). Besides, the mass field around
each particle in the system becomes more concentrate. This is obvious from
the difference between mass amplitudes written for a system's particle and a
free particle
\begin{equation}
\label{56} \left( {\frac{{l_{{\kern 1pt}\rm  Planck}}
}{{r}}{\kern 2pt}m_{{\kern 0.5pt} 0} \left| {\cos{\kern 1pt}
{\kern 1pt} \left( {K{\kern 1pt} r} \right){\kern 1pt}} \right| -
\frac{{l_{{\kern 1pt} \rm Planck}} }{{r}} {\kern 2pt} m_{{\kern
0.5pt} 0} \left| {\cos{\kern 1pt} {\kern 1pt} \left( {k{\kern
1pt} r} \right){\kern 1pt}}  \right|} \right)_{\,r {\kern 1pt}
\ll {\kern 1pt} \Lambda}  \;\,\, \propto \,\,{\kern 1pt}
\,m_{{\kern 1pt} 0} {\kern 1pt} \chi ^{2}{\kern 1pt}
\Theta^{2}/2{\kern 1pt} {\kern 1pt} r\,{\kern 1pt} \,\, > \,\,0.
\end{equation}

This is a typical phenomenon known as the defect of mass (especially in the
nuclear physics), though so far the phenomenon has not been studied in such
the broad sense.

Let us investigate how the system of particles will behave when external
conditions, namely, thermodynamic parameters, change. Evidently that any
alteration in the system's parameters affects proper particle parameters,
such as the value of the particle velocity $\upsilon $, the spatial
amplitude of the velocity oscillation (the de Broglie wavelength) $\lambda $
and the amplitude of the particle's inerton cloud $\Lambda $. The three
given parameters are concerned with the period of the particle oscillation
$\Theta$  [15-17]:

\begin{equation}
\label{57}
\Theta {\kern 1pt} = \, \Lambda /c\,\,\, =
\,\lambda /\upsilon.
\end{equation}

At the same time it is this characteristic $\Theta$ that enters
expression (56). This automatically means that any change in thermodynamic
parameters will immediately change $\upsilon ,\,\,{\kern 1pt} {\kern 1pt}
\lambda $ and $\Lambda $, which in turn must influence the value of $\Theta$
in the solution for the mass distribution (55). Therefore, the change
between two equilibrium states of thermodynamic parameters signifies the
appearance of the difference in the value of mass amplitude of the
particle's inerton cloud, which for states 1 and 2 can be written as (when
the inequality $K_{1} {\kern 1pt} r \ll 1$ holds)
\[
\frac{{m_{0}} }{{r}}\left( {\left| {\cos  \left( {K_{2}
{\kern 1pt} r} \right){\kern 1pt}}  \right|\, - \left| {\cos{\kern 1pt}
{\kern 1pt} \left( {K_{1} {\kern 1pt} r} \right){\kern 1pt}}  \right|}
\right)\,\,{\kern 1pt} \,
\]
\begin{equation}
\label{58}
 \approx \,\,\frac{{m_{0}} }{{r}}\left[ {\cos \left(
{k_{1} \sqrt {1 - \chi ^{2}\left( {\Theta_{1} + \delta \Theta} \right)^{2}}
{\kern 1pt} r} \right) - \cos  \left( {k_{1} \sqrt {1 -
\chi ^{2}\Theta_{1}^{2}}  {\kern 1pt} r} \right)} \right] \approx {\kern
1pt} {\kern 1pt} \,\frac{{m_{0}} }{{r}}{\kern 1pt} \chi ^{2}{\kern 2pt}
\Theta_{1} {\kern 1pt} {\kern 1pt} \delta {\kern 1pt} \Theta.
\end{equation}

Thus the value of mass, which is carried by inertons from the particle's
inerton cloud at the rearrangement of the system, can be crudely estimated
as $\delta m = m_{0} {\kern 1pt} \chi ^{2}{\kern 1pt} \Theta_{1} {\kern
1pt} {\kern 1pt} \delta {\kern 1pt} {\kern 1pt} \Theta$, or in the
explicit form
\begin{equation}
\label{59}
\delta {\kern 1pt} m = \chi^2 {\kern 1pt} h^2 {\kern 1pt} \delta {\kern
1pt} \upsilon {\kern 1pt} ( m_{{\kern 1pt} 0} {\kern 1pt} \upsilon^5);
\end{equation}
we have taken into account the de Broglie wavelength of the particle $\lambda =
h/( {m_{{\kern 1pt} 0} {\kern 1pt} \upsilon_0} )$, where $h$
is Planck's constant, and relations (57) that give $\Theta_1 h/(m_0v)^2$
and $\pm {\kern 1pt} \delta {\kern 1pt}\Theta = \mp {\kern 1pt} h{\kern 1pt}
\delta \upsilon /(m_{{\kern 0.5pt} 0} {\kern 1pt} \upsilon^3) $.
Then for the system of $N$ particles the value of the mass, which the system
emits or acquires at its thermodynamic rearrangement, will become

\begin{equation}
\label{60}
\Delta {\kern 1pt} m = \chi ^{2}{\kern 1pt} h^{2}{\kern 1pt} N\,\delta
{\kern 1pt} \upsilon \,/{\kern 1pt} \left( {m_{{\kern 1pt} 0} \upsilon ^{5}}
\right)
\end{equation}

Let us now come back to the example considered in section 3. If we divide
the mass $2.23 \times 10^{ - 4}$ kg of the water specimen by the mass of
water molecule $m_{0} \approx 3 \times 10^{ - 26}$ kg, we obtain the number
of molecules in the specimen $N \approx 0.74 \times 10^{22}$.

The value of ``quantum" velocity $\upsilon $ can be derived in the following
way. From the relationship $h\nu = m_{0} \upsilon _{0}^{2} /2$, where one
should put $\nu = 1/(2\Theta)$ [15-17], we obtain $\upsilon _{0} = 1490$
m/s at $\nu = 10^{14}$ s$^{-1}$ that is the typical value for water
molecules. The temperature introduces a disturbance to $\upsilon _{0} $,
which can be estimated from the relationship ${\textstyle{{3} \over
{2}}}k_{\rm B} T = {\textstyle{{1} \over {2}}}m_{0} {\kern 1pt} \upsilon
_{\rm th}^{2} $, namely, the thermal velocity $\upsilon _{\rm th}
= \sqrt {3k_{\rm B} T/m_{0}}  $; the difference between these characteristics exactly
corresponds to the correction $\delta \upsilon $ in expression (60).
Numerical values to this parameter are the following: $\upsilon
_{\rm th}^{\left( {\rm initial} \right)} = 64$ m/s at the normal (initial)
conditions and $\upsilon _{\rm th}^{\left( {\rm equilib} \right)} = 66$ m/s at the
equilibrium temperature 313 K. Therefore, the value of $\upsilon $ in
expression (60) becomes $\upsilon = \upsilon _{0} + \upsilon _{\rm th}^{\left(
{\rm initial} \right)} = 1554$ m/s and $\delta \upsilon = \upsilon _{\rm th}^{\left(
{\rm equilib} \right)} - {\kern 1pt} {\kern 1pt} {\kern 1pt} {\kern 1pt}
\upsilon _{\rm th}^{\left( {\rm initial} \right)} = 2$ m/s.

Although $\chi $ is a fitting parameter, we can evaluate it as well. Indeed,
in the Lagrangian (48) $\chi $ plays the role of a frequency at which
particles together with their inerton clouds exchange by local deformations
and tensions of space. In the system of identical particles, which is
treated, the frequency $1/\Theta$ can be called fundamental and hence
one can expect that $\chi \propto 1/\Theta$, though as it follows from
(55), $\chi  \ll  1/\Theta$. Since particles in the system are specified
by velocities $\upsilon $ and \textit{c}, we may suggest that the following
combination exactly represents the parameter $\chi $:
\begin{equation}
\label{61}
\chi = \upsilon /( {\Theta{\kern 1pt} c}).
\end{equation}

Setting for the velocity of inertons, carriers of quantum mechanical and
gravitational interactions, $c = 2.46 \times 10^{10}$ m/s, we obtain from
expression (60): $\Delta {\kern 1pt} m = 2.84 \times 10^{ - 17}$ kg. This
estimate checks well with the result (24) obtained in the framework of the
thermodynamic analysis stated above.

\section*{6. A system of particles in an external \\ dynamic mass field}

\subsection*{6.1. The behavior of the particle mass}

\hspace*{\parindent}
If the system of particles under consideration falls within an external mass
field, the Lagrangian (48) must be supplemented by one more term that
describes the interaction of the field with masses inside the system. What
can produce such mass field? In paper [20] this question has been discussed
in detail: A moving object induces its own inerton field in the ambient
space, which transfers mass, and hence a test material object (a particle,
or a system of particles) must undergoes the influence of this inerton
(mass) field. Evidently the term of interaction should be written by analogy
with the last term in the Lagrangian (48), namely, the given Lagrangian is
supplemented by the term
\begin{equation}
\label{62}
 - {\textstyle{{\pi}  \over {2}}}\sum\limits_{\vec n} {\chi _{\vec
{n}}^{\rm ext} {\kern 1pt}}  \dot {m}{\kern 1pt} _{\vec n} {\kern 1pt} {\vec
\xi} ^{{\kern 1pt}\rm ext}.
\end{equation}

Then instead of equation (52) we obtain the equation
\begin{equation}
\label{63}
\ddot {\mu} _{\vec {n}} - c^{2}\Delta {\kern 1pt} \mu _{\vec {n}} +
{\textstyle{{\pi}  \over {2}}}\chi ^{2}\left( {n} \right)\mu _
{\vec {n}} -
{\textstyle{{\pi}  \over {2}}}\chi _{\vec n}^{{\kern 1pt}\rm ext}
{\kern 1pt}\dot {\vec {\xi} }
\approx 0.
\end{equation}

The solution to equation (63) can be written as follows
\begin{eqnarray}
 m {\kern 1pt} _{\vec n} \left( r, {\kern 3pt} t \right)
 &=& \frac{{l_{\rm Planck}} }{{r}}{\kern 1pt}m_{0} {\kern 1pt} \left|
{{\kern 1pt} \cos\left( {K_{\vec {n}} r} \right)} \right| \left|
{\cos\left( {\Omega _{\vec {n}} {\kern 1pt} t}
\right)} \right|{\kern 1pt}    \nonumber    \\
 && + \left( { - 1} \right)^{\left[
t/\left( 2\pi /\Omega ^{\rm ext} \right ) \right]} {\kern 1pt}
  \frac{\chi ^{\rm ext} {\kern 1pt} \xi _0^{\rm ext}{\kern 1pt}
   \Omega ^{\rm ext}} {R^{\rm ext}} {\kern 1pt}\left| {{\kern 1pt}
    \sin\left( {K^{\rm ext}R^{\rm ext}}
\right)}  \right| \left|{\kern 0.5pt} {\cos\left(
{\Omega ^{\rm ext}{\kern 1pt} t} \right)} \right|;
\label{64}
\end{eqnarray}

\noindent
here is the mean distance from the external origin of the mass field to the
system studied and omitted is the mean frequency of oscillations
of the external mass field.

Thus expression (64) shows that particles in the system affected by an
external mass field become more weight.

\subsection*{6.2. A specific example, quantum mechanics and \\ simulation}

\hspace*{\parindent}
It is interesting to examine how the increase in the particle mass can be
treated in the framework of the orthodox quantum mechanical approach. Let us
simulate the behavior of the aqueous solution of hydrogen peroxide affected
by a technology developed in 1986, called Teslar, a tiny chip inserted into
a quartz wrist-watch; this Teslar chip has recently been studied by the
Krasnoholovets' research team [22].

The Teslar wristwatch developed 18 years ago is a special chip, which
neutralizes two currents (or cancels out two flows of electromagnetic
field), and generates a low frequency scalar wave due to this reaction of
neutralization. That was the hypothesis of the authors of the invention. The
hypothesis was tested in a number of ways in the late 1980s, including using
in-vitro analysis of rat nerve cells and human immune cells, in which
noradrenaline uptake into nerve cells was inhibited by as much as 19.5\%
more than in controls and lymphocyte proliferation increased by as much as
76\% over controls [23,24]. Additional reports - using electro-dermal
screening procedures, which measure the conductivity of acupuncture points,
- suggested that when a Teslar watch is worn on the left wrist there are
dramatic improvements in the energy level readings of most organs [25].
Because of such reports and the desire to further understand the physics of
the Teslar technology, additional scientific studies have been commissioned.

In Figure 1 we show typical infrared spectra recorded from aqueous solution
of hydrogen peroxide. The spectra demonstrate that the Teslar watch
suppresses the vibration of molecules in the solution under consideration
and, in particular, strongly freezes vibrations of O-H bonds, which is
especially seen in the vicinity of the maximum 3400 cm$^{-1}$. We suppose
that the Teslar chip can strongly affect a significantly non-equilibrium
system and the hydrogen bond is a good marker for the observation of changes
in the system studied.

From the viewpoint of the theory sated above we may conclude that the
solution acquires an additional mass radiated by the Teslar watch. However,
the means of conventional quantum mechanics do not allow any similar
explanation. What exactly can quantum mechanics propose?
\begin{figure}
\begin{center}
\includegraphics[scale=0.7]{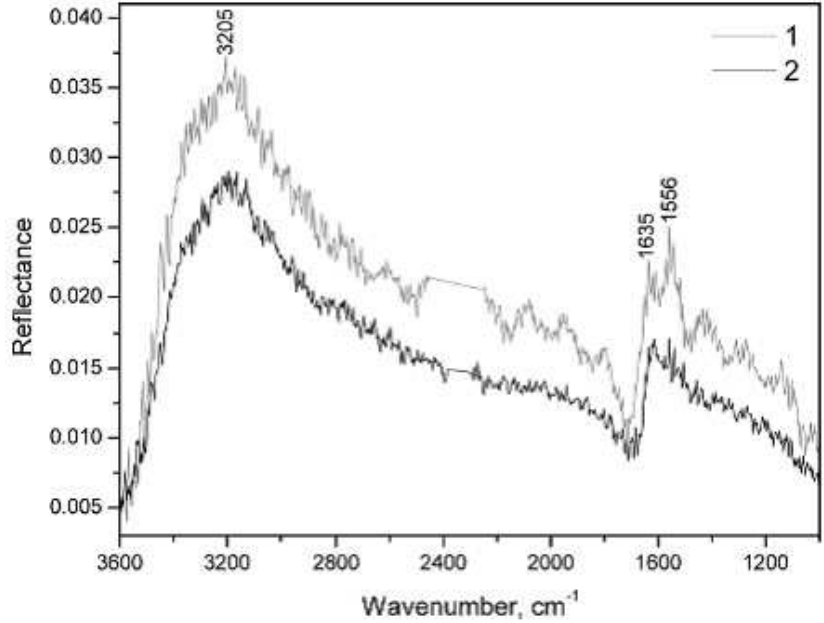}
\caption{\small IR-RAS spectra of H$_{2}$O$_{2}$ solution just after
pouring it in the cuvette. 1 (upper curve): the sample did not undergo the
influence of the Teslar watch; 2 (lower curve): the sample was affected by
the Teslar watch.} \label{Figure 1}
\end{center}
\end{figure}

The behavior of a quantum system obeys the Schr\"odinger equation, which in
our case can be written as follows
\begin{equation}
\label{65}
i{\kern 1pt} \hbar \frac{{\partial \psi \left( {x,t} \right)}}{{\partial
{\kern 1pt} t}} = - \frac{{\hbar ^{2}}}{{2m}}\frac{{d^{2}\psi \left( {x,t}
\right)}}{{d{\kern 1pt} x^{2}}} + V\left( {x,t} \right){\kern 1pt} {\kern
1pt} \psi \left( {x,t} \right)
\end{equation}

\noindent
and the initial wave function $\psi \left( {x,t} \right)$ is given as
\begin{equation}
\label{66}
\psi \left( {x,{\kern 1pt} {\kern 1pt} {\kern 1pt} 0} \right) = \left( {2\pi
\delta _{0}^{2}}  \right)^{ - 0.25} {\kern 1pt} \exp\left[ {i{\kern 0.5pt}
p_{{\kern 0.5pt}0} {\kern 0.5pt} x/\hbar - \left( {x - x_{0}}  \right)^
{2}/\left( {2\delta _{0}}  \right)^{2}}\right].
\end{equation}

Here $x_{0}$, $\delta _{{\kern 0.5pt}0}$ and $p_{{\kern 0.5pt} 0}$
are the initial position, width and momentum for the wave function of
hydrogen atom, respectively. Figure 2 shows the base state of that
hydrogen atom on O-H bonding.
\begin{figure}
\begin{center}
\includegraphics[scale=0.7]{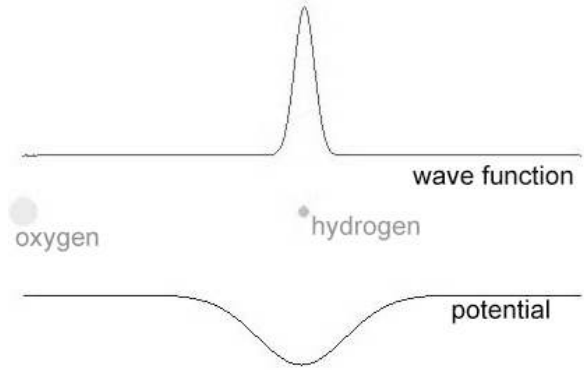}
\caption{\small The base state of hydrogen atom on the O-H bonding.}
\label{Figure 2}
\end{center}
\end{figure}

Figure 3 shows the feature of the absolute value of wave function of that
hydrogen atom after being irradiated by the infrared beam from FTIR.

The wave function of hydrogen atom vibrates right and left absorbing the
infrared energy. However, when turning on the Teslar watch, the wave is
broken in pieces and the vibration nearly stops due to the effect of inerton
absorption as shown in Figure 4. The absorption of inertons can be modulated
in quantum mechanics by introducing of the potential \textit{V} in equation
(65) in the form similar to the Newton's one,
\begin{equation}
\label{67}
V\left( {x,t} \right) = - \frac{\alpha } {x} {\kern 1pt} \cos\left( {\Omega {\kern 1pt}
{\kern 1pt} t} \right).
\end{equation}

\begin{figure}
\begin{center}
\includegraphics[scale=0.7]{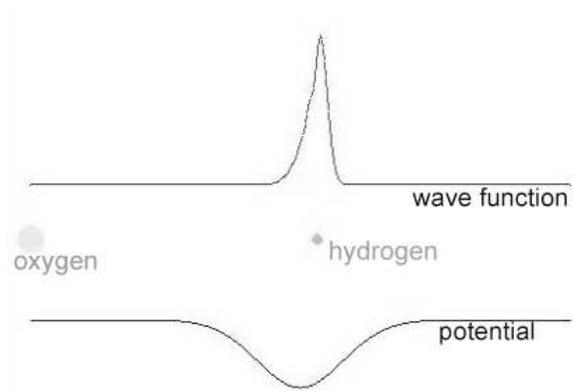}
\caption{\small The wave function of the hydrogen atom changes under the
IR from FTIR.}
\label{Figure 3}
\end{center}
\end{figure}

There have been the following choices and parameters that have been tuned in
our simulator: 1)\underline { Rep. No.} is the number of iterations of the
discretized Schr\"odinger equation; 2) \underline {IR.p} is the wave length
of irradiated IR (the resonant wave length is around 300); \underline
{I.pot} is the amplitude of additional potential brought by inertons. It
corresponds to the $\alpha $ in our equation with the multiplier: 50,000,
for example, the 5,000 for \underline {I.pot} is equivalent to 0.1 for
$\alpha $; 3) \underline {I.delta} is the coefficient that is used for
modifying the additional term in our equation (65) as
\[
\left( {\alpha /x} \right) \times \cos\left( {\Omega  {\kern 1pt}
t} \right) \Rightarrow \left( {\alpha /\left( {\rm x + I.delta} \right) \times
\cos\left( {\Omega {\kern 1pt} t} \right)} \right)
\]

\noindent
to avoid the singularity on $x = 0$; 4) \underline { J.pot (off/on)} allow us
to consider the effect of another O-H bonding. If  \underline {J.pot} is on,
the distance of these two O-H bonding is turned with the parameter ``another
.H"; 5) \underline {FTIR (off/on)} allow us to treat whether FTIR is on or
off.

\begin{figure}
\begin{center}
\includegraphics[scale=0.7]{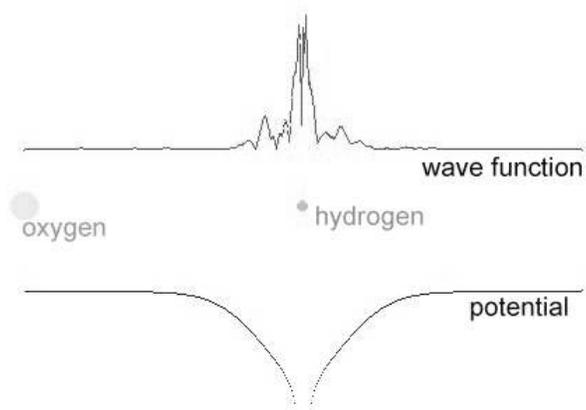}
\caption{ \small The wave is broken in pieces and the vibration nearly
stops due to the effect of inerton absorption when turning on the Teslar
watch.} \label{Figure 4}
\end{center}
\end{figure}

\bigskip

\begin{figure}
\begin{center}
\includegraphics[scale=0.7]{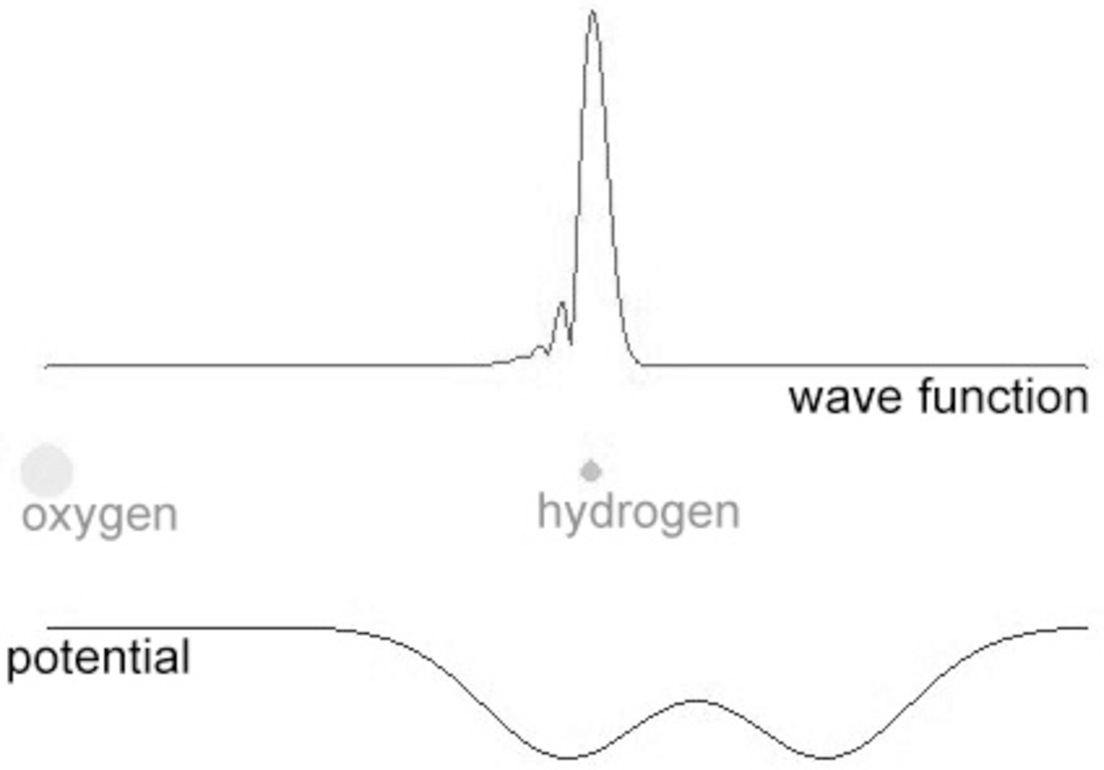}
\caption{ \small Hydrogen bonding between two water molecules.}
\label{Figure 5}
\end{center}
\end{figure}

As another possibility, we may try to simulate the effect of hydrogen
bonding between two water molecules. This time the potential has two valleys
caused from two different O-H bonding as shown in Figure 5. In this case the
feature of vibration of hydrogen atom is almost the same as the case without
considering the other O-H bonding shown in Figure 3. However when reducing
the distance between two valleys, i.e. making closer two water molecules,
the wave has broken in pieces and the vibration of wave function of hydrogen
atom has ceased as shown in Figure 6. This means that O-H bonding no longer
absorbs the energy from IR.

Thus the simulation of quantum mechanical problem by a classical scheme
gives an information only on the behavior of $\psi $ function and does not
clarify the reasons for the potential \textit{V} that remains pure classical
in non-deterministic quantum equation (65). The submicroscopic concept
allows us to account for the origin of the potential \textit{V} and gives a
deeper analysis of the problem.

\begin{figure}
\begin{center}
\includegraphics[scale=0.7]{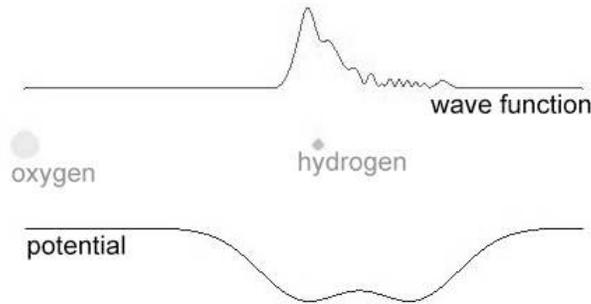}
\caption{ \small  Two water molecules become closer; the wave function has
broken in pieces and the vibration of wave function of hydrogen atom has ceased.}
\label{Figure 6}
\end{center}
\end{figure}

\section*{7. Concluding remarks}

\hspace*{\parindent}
As many scientists who have dealt with thermodynamics, we have often had the
impression that thermodynamics is easier to use than to understand. In such
a situation, it is of real comfort to discover in books written by other
scientists that they have felt the same impression. While two of these books
only have been referenced herein, Refs. [3,4], we would like to express our
gratitude to all the authors who have transmitted such a message, which is
indubitably of great scientific interest.

In the present paper, by means of both conventional thermodynamic methods
and the submicroscopic analysis, we have shown that the change in mass is a
general law of evolution of physical systems. The mass exchange is the
fundamental property of any physical system, from a system of several
canonical particles to that of cosmological objects. Hitherto the defect of
mass has being taken into account only in problems associated with nuclear
physics. However, the present research points to the necessity of including
the phenomenon of the defect of mass also in tasks typical for condensed
matter physics. Besides, the submicroscopic consideration has allowed us to
give an estimation of the speed of inertons, carriers of quantum mechanical
and gravitational interactions of objects, which has been evaluated as about
$2.5 \times 10^{10}$ m/s.

\bigskip

\subsection*{Acknowledgement}

\hspace*{\parindent}
We thank greatly Harry Yosh for the help in simulation.

\end{document}